\documentclass[prd,twocolumn,showpacs,floatfix,amsmath,amssymb,floatfix]{revtex4}
\usepackage{graphicx,color,dcolumn,booktabs,bm}
\usepackage{longtable,lscape}
\usepackage{txfonts}
\usepackage{overpic}
\usepackage{amssymb}
\usepackage{indentfirst}
\usepackage{feynmf}   
\usepackage{slashed}  
\usepackage{cases}
\usepackage{color}
\usepackage{multirow}
\usepackage{epstopdf}
\usepackage{graphicx,color,dcolumn,booktabs,bm}

\graphicspath{{Figures/}} %
\usepackage[colorlinks,
            citecolor=blue,
            anchorcolor=red,
            menucolor=red,
            linkcolor=red,
            filecolor=red,
            runcolor=red,
            urlcolor=blue,
            frenchlinks=red]{hyperref}
\begin{document}

\title{hidden charm decays of $X(4014)$ in a $D^{*}\bar{D}^{*}$ molecule scenario}

\author{Zi-Li Yue$^{1}$}
\author{Man-Yu Duan$^{1}$}
\author{Chun-Hua Liu$^{1}$}\email{liuch@seu.edu.cn}
\author{Dian-Yong Chen$^{1,2}$}\email{chendy@seu.edu.cn}
\author{Yu-Bing Dong$^{3,4}$}\email{dongyb@ihep.ac.cn}
\affiliation{
 $^{1}$ School of Physics, Southeast University,  Nanjing 210094, China\\
 $^2$Lanzhou Center for Theoretical Physics, Lanzhou University, Lanzhou 730000, China\\
$^3$Institute of High Energy Physics, Chinese Academy of Sciences, Beijing 100049, China\\
$^4$School of Physical Sciences, University of Chinese Academy of Sciences, Beijing 101408, China}

\begin{abstract}
Inspired by the recent observation of a new structure, $X(4014)$, in the process $\gamma\gamma\to \gamma\psi(2S)$, we evaluate the possibility of assigning $X(4014)$ as a $D^\ast \bar{D}^\ast$ molecular state with $I(J^{PC})=0(0^{++})$ by investigating the hidden charm decays of $X(4014)$. The partial widths of $J/\psi\omega$, $ \eta_{c}\eta$ and $\eta_{c}\eta^{\prime}$ channels are evaluated to be about $(0.41\sim 5.00)$, $(2.05\sim7.49)$ and $(0.11\sim0.51)\ \mathrm{MeV}$, respectively. Considering the experimental observation and the present estimations, we proposed  to search $X(4014)$ in the $\gamma \gamma \to J/\psi \omega$ process in Belle II. 
\end{abstract}


\maketitle

\section{Introduction}
\label{sec:introduction}

Similar to the deuteron composed of a proton and a neutron, the deuteron-like molecule states~\cite{Weinberg:1965zz} composed of other hadrons are expected to exist~\cite{Xiao:2016hoa,Dong:2017gaw,Faessler:2007gv,Faessler:2007us,Wang:2020dgr,Lebed:2022vks}. Searching for such kind of molecule states experimentally is an intriguing topic in hadron physics. The first promising candidate is the long standing and well established $X(3872)$, which was observed by 
 the  Belle Collaboration in the year of 2003~\cite{Belle:2003nnu} , and the observed mass is very close to the threshold of $D^\ast \bar{D}$~\cite{Liu:2008fh,Wu:2021udi,Voloshin:1976ap,DeRujula:1976zlg,Tornqvist:1993ng,Tornqvist:2004qy,Wang:2020dgr}. The $I(J^{PC})$ quantum numbers have been determined to be $0(1^{++})$. Besides the $X(3872)$, there exist another two states near the threshold of $D^\ast \bar{D}/D^\ast D$, which are $Z_c(3900)$~\cite{Wu:2019vbk,Dong:2013iqa,Liu:2009qhy,Sun:2012zzd}  and $T_{cc}(3875)$~\cite{Li:2012ss,Xu:2017tsr,Liu:2019stu,Ding:2020dio,Li:2021zbw,Chen:2021vhg,Agaev:2021vur,Ren:2021dsi,Albaladejo:2021vln,Dong:2021bvy,Baru:2021ldu,Du:2021zzh}, respectively. The former state, $Z_c(3900)$ was firstly observed in 2013 by the BESIII~\cite{BESIII:2013ris} and Belle~\cite{Belle:2013yex} Collaboration in the $\pi^\pm J/\psi$ invariant mass spectrum of the process $e^+ e^- \to \pi^+ \pi^- J/\psi$ at $\sqrt{s}=4260$ MeV, and later, the authors in Ref.~\cite{Xiao:2013iha} confirmed the existence of $Z_c(3900)$ by using the data sample collected by CLEO-c Collaboration at $\sqrt{s}=4170$ MeV. The $I(J^P)$ quantum numbers of $Z_c^0(3900)$ are determined to be $1(1^{+-})$. As for the later state $T_{cc}(3875)$, it was observed by the LHCb Collaboration in the $D^0 D^0 \pi^+$ invariant mass spectrum in 2021 \cite{LHCb:2021auc,LHCb:2021vvq}. Since the measured masses of $X(3872)$, $Z_c(3900)$ and $T_{cc}(3875)$ are all close to the threshold of $D^\ast \bar{D}/D^\ast D$, these states have been extensively investigated in the molecular scenario~\cite{Xiao:2016hoa,Dong:2017gaw,Faessler:2007gv,Faessler:2007us, Lebed:2022vks,Wang:2020dgr}. If $X(3872)$, $Z_c(3900)$ and $T_{cc}(3875)$ are all molecular, it seems that the interactions between $S$ wave charmed meson are attractive and strong enough in various channels.

Besides the states near the threshold of $D^\ast \bar{D}$, there is another state named $Z_c(4020)$~\cite{He:2013nwa,Chen:2013omd} near the threshold of $D^\ast \bar{D}^\ast$, which could be considered as deuteron-like molecular state composed of $D^\ast \bar{D}^\ast$. The state $Z_c(4020)$ was firstly observed in the $\pi^\pm h_c$ invariant mass spectrum of the process $e^+ e^- \to \pi^+\pi^- h_c$ by the BESIII Collaboration at $\sqrt{s}=4260$ MeV in 2013~\cite{BESIII:2013ouc}. Extensive investigations from different aspects, such as mass spectrum~\cite{Wang:2022fdu,Sakai:2021qrg,Ortega:2018cnm}, decay~\cite{Wang:2022fdu,Sakai:2021qrg,Xiao:2019spy,Ke:2016owt} and production properties~\cite{Wu:2019vbk} from various groups indicated that $Z_c(4020)$ could be assigned as a $D^\ast \bar{D}^\ast$ molecular state. Similar to the case of the prosperous states near the $D^\ast \bar{D}$ threshold, it is expected that there also exist abundant molecular candidates near the threshold of $D^\ast \bar{D}^\ast$.

Recently, the Belle Collaboration reported their measurements of the cross section for the two-photon process $\gamma \gamma \to \gamma \psi(2S)$ from the threshold to 4.2 GeV~\cite{Belle:2021nuv}. Two structures were observed in the cross sections, and the one with a mass of $3922.4 \pm 6.5 \pm 2.0$ MeV and a width of  $22 \pm 17 \pm 4$ MeV could be considered $X(3915)$, $\chi_{c2}(3930)$ or an admixture of them. Besides there exists a new state at 4014 MeV, hereinafter, we named this new state as $X(4014)$. The mass and width of $X(4014)$ were reported to be, 
\begin{eqnarray}
M &=& (4014.3 \pm 4.0\pm 1.5)\ \mathrm{MeV}, \nonumber \\
\Gamma&=& (4\pm 11\pm 6) \ \mathrm{MeV}, 
\end{eqnarray}
respectively. 

It is interesting to notice that the newly observed $X(4014)$ matches none of the known charmonium or charmonium-like states, and moreover, its mass is just several MeV below the threshold of $D^\ast \bar{D}^\ast$, which is similar to the case of  $Z_c(4020)$. Thus, one can consider $X(4014)$ as a candidate of $D^\ast \bar{D}^\ast$ molecular state with positive $C$-parity. In Ref. \cite{Duan:2022upr}, we investigate the molecular possibilities of $X(4014)$ in the framework of the local hidden gauge approach~\cite{Liu:2005jb}, and we find the most possible $I(J^{PC})$ numbers of $X(4014)$ are $0(0^{++})$. In the present work, we further test the molecular possibility of $X(4014)$ by investigating the hidden charm decay properties of $X(4014)$ in such a $D^\ast \bar{D}^\ast$ molecular scenario with effective Lagrangian approaches~\cite{Wu:2021ezz,Wu:2021cyc,Chen:2016ncs}. The predicted decay channels can be experimental accessible in by Belle II, which could be a crucial test of the $D^\ast \bar{D}^\ast$ molecular assignment of $X(4014)$.

This work is organized as follows. The hadronic molecular structures of $X(4014)$ is discussed in the following section and the hidden charm decays processes, including $X(4014)\to J/\psi \omega$, $ \eta_{c}\eta $ and $ \eta_c \eta^{\prime}$ are estimated in Sec.\ref{sec:Sec3}. The numerical results and the relevant discussions are presented in \ref{sec:Sec4}, and the last section is dedicated to a short summary.

\section{Hadronic molecular structure}
\label{sec:Sec2}

In the present work, the newly observed $X(4014)$ is assigned as a molecular state compose of $D^\ast \bar{D}^\ast$ with $I(J^{PC})=0(0^{++})$. The interaction between $X(4014)$ and its components can be described by an effective Lagrangian in the form,
\begin{eqnarray}
\mathcal{L}_{X}&=&\frac{g_{X}}{\sqrt{2}} X\int dy\Phi_{X}(y^{2})\Bigg[D_{\mu}^{*+}\Big(x-\frac{y}{2}\Big)D^{*-\mu}\Big(x+\frac{y}{2}\Big)\nonumber\\&+&D^{*0}_{\mu}\Big(x-\frac{y}{2}\Big)\bar{D}^{*0\mu} \Big(x+\frac{y}{2}\Big)\Bigg],\label{Eq:LagX}
\end{eqnarray}
where $\Phi_X(y^2)$ is the correlation function, which is introduced to describe the distribution of the $D^{*}$ and $\bar{D}^{*}$ meson in the molecular state.

The Fourier transformation of the correlation function is
\begin{eqnarray}
\Phi_{X}(y^{2})=\int \frac{d^{4}p}{(2\pi)^{4}}e^{-ipy}\tilde{\Phi}_{X}(-p^{2}).
\end{eqnarray}
An appropriate $\tilde{\Phi}_{X}(-p^{2})$ should not only describe the interior structure of the molecular state, but also fall fast enough in the ultraviolet region. Here, we employ the form factor in the Gaussian form~\cite{Faessler:2007gv,Faessler:2007us}, 
\begin{eqnarray}
\tilde{\Phi}_{X}(p_{E}^{2})=\mathrm{exp}(-p_{E}^{2}/\Lambda^{2})
\end{eqnarray}
where $\Lambda$ is a model parameter for parametrizing the distribution of the components inside the molecular state. 

\begin{figure}[t]
  \centering
  \includegraphics[width=6.3cm]{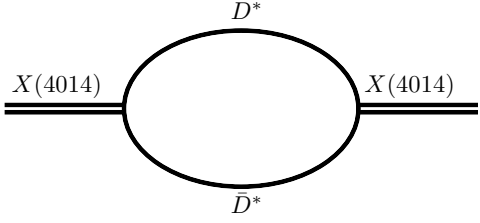}
  \caption{The mass operator of the $X(4014)$. Here $D^\ast$ and $\bar{D}^\ast$ refer to $(D^{\ast +}, D^{\ast 0})$ and $(D^{\ast -}, \bar{D}^{\ast 0})$, respectively}\label{Fig:Tri1}
\end{figure}

As a composite particle, the renormalization constant of the $X(4104)$ should be zero~\cite{Hayashi:1967bjx, Salam:1962ap, Weinberg:1962hj}, which can be used to determine the coupling constant between the molecular and its components, i.e.,
\begin{eqnarray}
\label{eq:1}
Z=1-\Pi^{\prime}(m^{2}_{X})=0, \label{Eq:ComZ}
\end{eqnarray}
with $\Pi^{\prime}(m^{2}_{X})$ is the derivative of the mass operator of the $X(4014)$, and the concrete forms of the mass operator of the $X(4014)$ corresponding to Fig. \ref{Fig:Tri1} is,
\begin{eqnarray}
\Pi ({m_{X}}^{2})&=&\int \frac{d^{4}q}{(2\pi)^{4}}\tilde{\Phi^{2}}[-(q-\frac{1}{2}p)^{2},\Lambda^{2}]\nonumber\\&\times&\frac{-g^{\mu\nu}+q^{\mu}q^{\nu}/m_{D^{*}}^{2}}{q^{2}-m_{D^{*}}^{2}}\nonumber\\&\times&\frac{-g^{\mu\nu}+(p-q)^{\mu}(p-q)^{\nu}/m_{\bar{D}^{*}}^{2}}{(p-q)^{2}-m_{\bar{D}^{*}}^{2}}
\end{eqnarray}

\section{Hidden charm decay}
\label{sec:Sec3}
\begin{figure}[htb]
\begin{tabular}{cc}
  \centering
  \includegraphics[width=4.2cm]{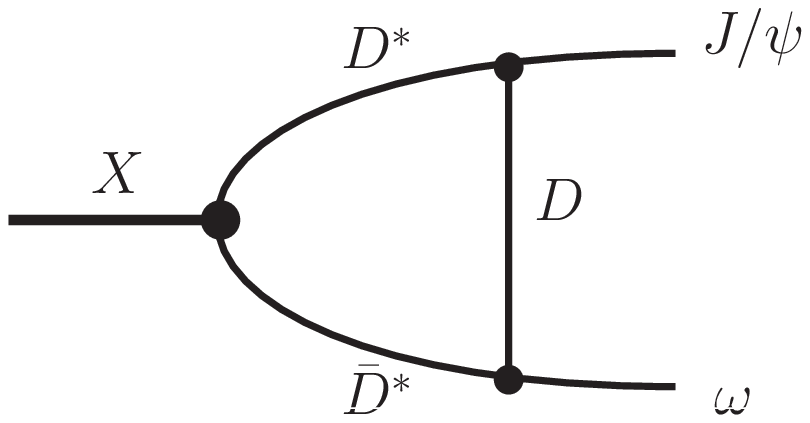}&
 \includegraphics[width=4.2cm]{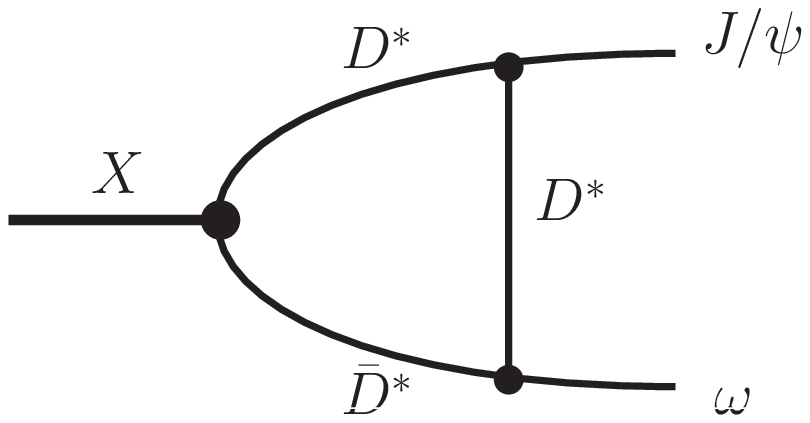}\\
\\
 $(a)$ & $(b)$ \\
  \includegraphics[width=4.2cm]{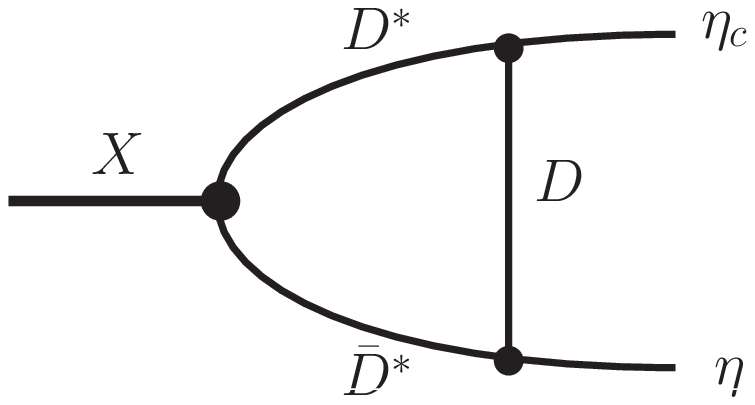}&
 \includegraphics[width=4.2cm]{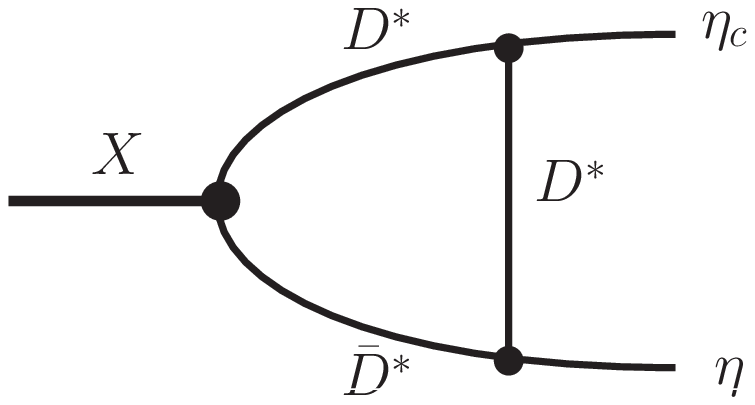}\\
\\$(c)$ & $(d)$ \\
  \includegraphics[width=4.2cm]{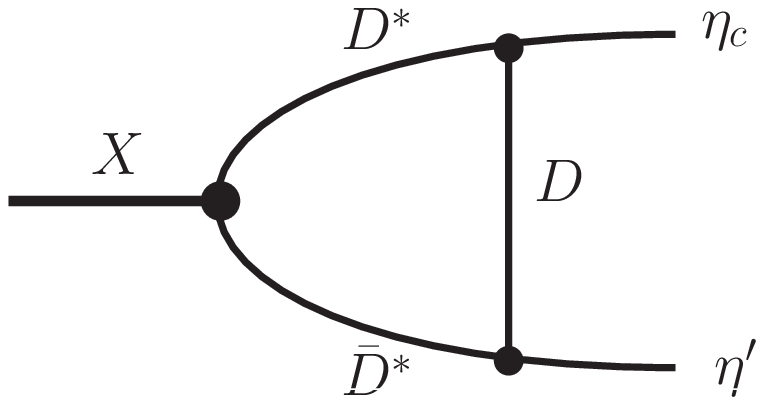}&
 \includegraphics[width=4.2cm]{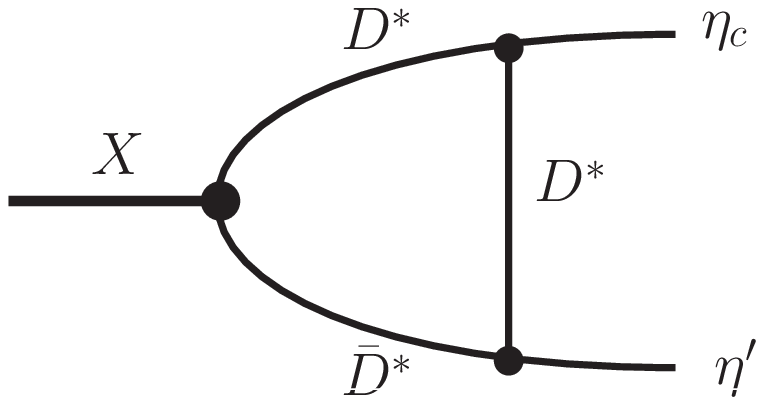}\\
 \\ $(e)$ & $(f)$ \\
 \end{tabular}
  \caption{The typical diagrams contributing to 
$X(4014)\rightarrow J/\psi\omega$ (diagrams (a) and (b)), $ X(4014)\rightarrow \eta_{c}\eta$ (diagrams (c) and (d)) and~$X(4014)\rightarrow \eta_{c}\eta^{\prime}$ (diagrams (e) and (f)).}\label{Fig:Tri}
\end{figure}

In the present work, we estimate the hidden charm decays of $X(4014)$ in the $D^\ast \bar{D}^\ast$ molecular scenario, where the $I(J^{PC})$ quantum numbers of the $X(4014)$ are considered as $0(0^{++})$. The possible decay channels include $X(4014) \to J/\psi \omega,\ \eta_c \eta,\ \eta_c \eta^\prime$. These decay processes are estimated in hadronic level, and the possible diagrams contributing to these processes are collected in Fig.~\ref{Fig:Tri}.

\subsection{Effective Lagrangians}
To estimate the diagrams in Fig.~\ref{Fig:Tri}, we employ an effective Lagrangian approach. The coupling between the molecular state and its components has been given in Eq.~(\ref{Eq:LagX}). Considering the heavy quark limit, one can construct the coupling between charmonia and the charmed meson pair, and the relevant effective Lagrangians are~\cite{Casalbuoni:1996pg, Oh:2000qr, Colangelo:2002mj, Neubert:2005mu},
\label{Sec:Num}
\begin{eqnarray}
\mathcal{L}_{\psi \mathcal{D}^{(*)}\mathcal{D}^{(*)}}&=&-ig_{\psi \mathcal{D}\mathcal{D}}\psi_{\mu}\mathcal{D}^{\dagger}\overleftrightarrow{\partial^{\mu}}\mathcal{D}\nonumber\\&+&g_{\psi D^{*}D}\epsilon^{\mu\nu\alpha\beta}\partial_{\mu}\psi_{\nu}(\mathcal{D}_{\alpha}^{*}\overleftrightarrow{\partial_{\beta}}\mathcal{D}^{\dagger}-\mathcal{D}\overleftrightarrow{\partial_{\beta}}\mathcal{D}_{\alpha}^{*\dagger})\nonumber\\&+&ig_{\psi \mathcal{D}^{*}\mathcal{D}^{*}}\psi^{\mu}(\mathcal{D}_{\nu}^{*}\overleftrightarrow{\partial^{\nu}}\mathcal{D}_{\mu}^{*\dagger}+\mathcal{D}_{\mu}^{*}\overleftrightarrow{\partial^{\nu}}\mathcal{D}_{\nu}^{*\dagger}\nonumber\\&-&\mathcal{D}_{\nu}^{*}\overleftrightarrow{\partial_{\mu}}\mathcal{D}^{*\nu \dagger}),\nonumber\\
\mathcal{L}_{\eta_{c}\mathcal{D}^{*}\mathcal{D}^{(*)}}&=&-ig_{\eta_{c}\mathcal{D}^{*}\mathcal{D}}\eta_{c}(\mathcal{D}\overleftrightarrow{\partial_{\mu}}\mathcal{D}^{*\dagger\mu}+D^{*\mu}\overleftrightarrow{\partial_{\mu}}\mathcal{D}^{\dagger})\nonumber\\&+&-g_{\eta_{c}\mathcal{D}^{*}\mathcal{D}^{*}}\epsilon^{\mu\nu\alpha\beta}\partial_{\mu}\eta_{c}\mathcal{D}^{*}_{\nu}\overleftrightarrow{\partial_{\alpha}}\mathcal{D}_{\beta}^{*\dagger},
\end{eqnarray}
where the $\mathcal{D}^{(*)\dagger}=(\bar{D}^{(*)0},D^{(*)-},D_{s}^{(*)-})$ and $A\overleftrightarrow{\partial_\mu}B=A\partial_\mu B-B\partial_\mu A$.

Considering the heavy quark limit and chiral symmetry, one can construct the effective coupling between light mesons and charmed meson pair, which is~\cite{Kaplan:2005es, Kaymakcalan:1983qq, Oh:2000qr, Casalbuoni:1996pg, Colangelo:2002mj}
\begin{eqnarray}
\mathcal{L}_{\mathcal{D}^{(*)}\mathcal{D}^{(*)}\mathcal{P}}&=&-ig_{\mathcal{D}^{*}\mathcal{D}\mathcal{P}}(\mathcal{D}^{i}\partial^{\mu}\mathcal{P}_{ij}\mathcal{D}_{\mu}^{*j\dagger}-\mathcal{D}_{\mu}^{*i}\partial_{\mu}\mathcal{P}_{ij}\mathcal{D}^{j\dagger})\nonumber\\&+&\frac{1}{2}g_{\mathcal{D}^{*}\mathcal{D}^{*}\mathcal{P}} \epsilon_{\mu\nu\alpha\beta}\mathcal{D}^{*\mu}_{i}\partial^{\nu}\mathcal{P}^{ij}\overleftrightarrow{\partial^{\alpha}}\mathcal{D}_{j}^{*\beta \dagger},\nonumber\\
\mathcal{L}_{\mathcal{D}^{(*)}\mathcal{D}^{(*)}\mathcal{V}}&=&-ig_{\mathcal{D}\mathcal{D}\mathcal{V}}\mathcal{D}_{i}^{\dagger}\overleftrightarrow{\partial_{\mu}}\mathcal{D}^{j}(\mathcal{V}^{\mu})^{i}_{j}\nonumber\\&-&2f_{\mathcal{D}^{*}\mathcal{D}\mathcal{V}}\epsilon_{\mu\nu\alpha\beta}(\partial^{\mu}\mathcal{V}^{\nu})^{i}_{j}(\mathcal{D}_{i}^{\dagger}\overleftrightarrow{\partial^{\alpha}}\mathcal{D}^{*\beta \dagger}-\mathcal{D}_{i}^{*\beta\dagger}\overleftrightarrow{\partial^{\alpha}}\mathcal{D}^{j})\nonumber\\&+&ig_{\mathcal{D}^{*}\mathcal{D}^{*}\mathcal{V}}\mathcal{D}_{i}^{*\nu\dagger}\overleftrightarrow{\partial_{\mu}}\mathcal{D}^{*j}_{\nu}(\mathcal{V}^{\mu})^{i}_{j}\nonumber\\&+&4if_{\mathcal{D}^{*}\mathcal{D}^{*}\mathcal{V}}\mathcal{D}^{*\dagger}_{i\mu}(\partial^{\mu}\mathcal{V}^{\nu}-\partial^{\nu}\mathcal{V}^{\mu})^{i}_{j}\mathcal{D}^{*j}_{\nu},
\end{eqnarray}
where $\mathcal{V}$ and $\mathcal{P}$ are the matrices form of vector nonet and pseudo-scalar nonet, and their concrete forms are,
\begin{eqnarray}
\mathcal{V}=
\begin{pmatrix}
\frac{1}{\sqrt2}(\rho^{0}+\omega)&\rho^{+}&K^{*+}\\
\rho^{-}&
\frac{1}{\sqrt2}(-\rho^{0}+\omega)&K^{*0}\\
K^{*-}&\bar{K}^{*0}&\phi\\ \nonumber
\end{pmatrix}
\end{eqnarray}
\begin{eqnarray}
\mathcal{P}=
\begin{pmatrix}
\frac{\pi^{0}}{\sqrt{2}}+\alpha\eta+\beta\eta^{\prime}&\pi^{+}&K^{+}\\
\pi^{-}&-\frac{\pi^{0}}{\sqrt{2}}+\alpha\eta+\beta\eta^{\prime}&K^{0}\\
K^{-}&\bar{K}^{0}&\gamma\eta+\delta\eta^{\prime}\\
\end{pmatrix},
\end{eqnarray}
where $\alpha$, $\beta$ and~$\delta$ are the parameters related to the mixing angle by. 
\begin{eqnarray}
\alpha &=&\frac{\mathrm{cos}\theta-\sqrt{2}\mathrm{sin}\theta}{\sqrt{2}}, \  \ \ \ \ \beta=\frac{\mathrm{sin}\theta+\sqrt{2}\mathrm{cos}\theta}{\sqrt{6}},\nonumber\\ \gamma &=&\frac{-2\mathrm{cos}\theta-\sqrt{2}\mathrm{sin}\theta}{\sqrt{6}}, \ \ \ \ \delta=\frac{-2\mathrm{sin}\theta+\sqrt{2}\mathrm{cos}\theta}{\sqrt{6}	}.\ \ \
\end{eqnarray}
where the mixing angle $\theta$ is determined to be $19.1^{\circ}$~\cite{MARK-III:1988crp, DM2:1988bfq}.

\subsection{Decay Amplitude}
With the effective Lagrangians listed above, we can get the amplitudes for $X(4014)(p_0) \to [D^{\ast}(p_1)\bar{D}^\ast(p_2)] D^{(\ast)}(q)\to J/\psi(p_3)\omega (p_4)$ corresponding to diagrams Fig.~\ref{Fig:Tri}-(a) and (b), which are,
\begin{eqnarray}
i\mathcal{M}_{a}&=&i^{3}\int \frac{d^{4}q}{2\pi^{4}}[g_{X}\tilde{\Phi}_{X}(-p_{12}^{2},\Lambda^{2}) g ^{\phi\tau}]\nonumber\\&\times&[g_{D^{*}D\psi}\epsilon_{\mu\nu\alpha\beta}(ip_{3})^{\mu}(iq+ip_{1})^{\beta}\epsilon_{\theta}(p_{3})g^{\nu\theta}g^{\alpha\zeta}]\nonumber\\&\times&[-2f\epsilon_{\sigma\omega\delta \lambda}(ip_{4})^{\sigma}\epsilon_{\rho}(p_{4})(-ip_{2}+q)^{\delta}g^{\omega\rho}g^{\lambda\iota}]\nonumber\\&\times&\frac{-g^{\phi x}+p_{1}^{\phi}p_{1}^{x}/m_{1}^{2}}{p_{1}^{2}-m_{1}^{2}}\frac{-g^{\tau a}+p_{2}^{\tau}p_{2}^{a}/m_{2}^{2}}{p_{2}^{2}-m_{2}^{2}}\frac{1}{q^{2}-m_{q}^{2}},\nonumber
\end{eqnarray}
\begin{eqnarray}
i\mathcal{M}_{b}&=&i^{3}\int \frac{d^{4}q}{2\pi^{4}}[g_{X}\tilde{\Phi}_{X}(-p_{12}^{2},\Lambda^{2}) g ^{\phi\tau}]\nonumber\\&\times&[ig_{D^{*}D^{*}\psi}g^{\mu\theta}(g^{\mu\sigma}g^{\nu\iota}(-ip_{1}-iq)^{\nu}+g^{\nu\zeta}g^{\mu\iota}(-ip_{1}-iq)^{\nu}\nonumber\\&-&g^{\nu\zeta}g^{\nu\iota}(-ip_{1}-iq)^{\mu})\epsilon_{\theta}(p_{3})]\nonumber\\&\times&[ig_{D^{*}D^{*}V}(-ip_{2}+q)^{\rho}g^{\omega\iota}g^{\omega\kappa}\epsilon_{\rho}(p_{4})+4if_{D^{*}D^{*}V}\nonumber\\&\times&(ip_{4}^{\sigma}g^{\omega\rho}-ip_{4}^{\omega}g^{\sigma\rho})g^{\sigma o}g^{\omega\iota}\epsilon_{\rho}(p_{4})]\nonumber\\&\times&\frac{-g^{\phi x}+p_{1}^{\phi}p_{1}^{x}/m_{1}^{2}}{p_{1}^{2}-m_{1}^{2}}\frac{-g^{\tau a}+p_{2}^{\tau}p_{2}^{a}/m_{2}^{2}}{p_{2}^{2}-m_{2}^{2}}\nonumber\\&\times&\frac{-g^{yb}+q^{y}q^{b}/m_{q}^{2}}{q^{2}-m_{q}^{2}},
\end{eqnarray}
where $p_{12}=(p_{1}-p_{2})/2$. $\epsilon(p_3)$ and $\epsilon(p_4)$ are the polarization vectors of $J/\psi$ and $\omega$ mesons, respectively.

As for the decay process $X(4014)\rightarrow \eta_{c}\eta$,  the amplitudes corresponding to diagrams Fig.~\ref{Fig:Tri}-(c) and (d) are, 
\begin{eqnarray}
i\mathcal{M}_{c}&=&i^{3}\int \frac{d^{4}q}{2\pi^{4}}[g_{X}\tilde{\Phi}_{X}(-p_{12}^{2},\Lambda^{2}) g ^{\phi\tau}]\nonumber\\&\times&[-ig_{D^{*}D\eta_{c}}(ip_{1}-q)^{\mu}g^{\mu\alpha}]\nonumber\\&\times&[-ig_{D^{*}D}p(ip_{4})^{\sigma}g^{\sigma\iota}]\frac{-g^{\phi x}+p_{1}^{\phi}p_{1}^{x}/m_{1}^{2}}{p_{1}^{2}-m_{1}^{2}}\nonumber\\&\times&\frac{-g^{\tau a}+p_{2}^{\tau}p_{2}^{a}/m_{2}^{2}}{p_{2}^{2}-m_{2}^{2}}\frac{1}{q^{2}-m_{q}^{2}},\nonumber\\
i\mathcal{M}_{d}&=&i^{3}\int \frac{d^{4}q}{2\pi^{4}}[g_{X}\tilde{\Phi}_{X}(-p_{12}^{2},\Lambda^{2}) g ^{\phi\tau}]\nonumber\\&\times&[-g_{D^{*}D^{*}\eta_{c}}\epsilon_{\mu\nu\alpha\beta}(ip_{3})^{\mu}\nonumber\\&\times&(-ip_{1}-q)^{\alpha}][\frac{1}{2}g_{D^{*}D^{*}V}\epsilon_{\sigma\omega\delta\lambda}(ip_{4})^{\omega}(-iq+ip_{2})^{\delta}\nonumber\\&\times&g^{\lambda\kappa}g^{\sigma\kappa}]\frac{-g^{\phi x}+p_{1}^{\phi}p_{1}^{x}/m_{1}^{2}}{p_{1}^{2}-m_{1}^{2}}\frac{-g^{\tau a}+p_{2}^{\tau}p_{2}^{a}/m_{2}^{2}}{p_{2}^{2}-m_{2}^{2}}\nonumber\\&\times&\frac{-g^{yb}+q^{y}q^{b}/m_{q}^{2}}{q^{2}-m_{q}^{2}}.
\end{eqnarray}

The amplitudes of the decay progress $X(4014)\rightarrow \eta_c \eta^{\prime}$ can be obtained by replacing the mass and relevant coupling constants of $\eta$ meson with those of $\eta^\prime$, i,e.,
\begin{eqnarray}
\mathcal{M}_{e}&=&\left.\mathcal{M}_{c}\right|_{\eta\rightarrow\eta^{\prime}},\\ \nonumber
\mathcal{M}_{f}&=&\left.\mathcal{M}_{d}\right|_{\eta\rightarrow\eta^{\prime}}.
\end{eqnarray}

The total amplitudes of $X(4014) \to J/\psi\omega, \ \eta_{c}\eta, \ \eta_{c}\eta^{\prime}$ are,
\begin{eqnarray}
\label{eq:16}
\mathcal{M}_{X\rightarrow J/\psi\omega}&=&\mathcal{M}_{a}+\mathcal{M}_{b},\nonumber \\ 
\mathcal{M}_{X\rightarrow J/\psi\eta}&=&\mathcal{M}_{c}+\mathcal{M}_{d}, \nonumber\\
\mathcal{M}_{X\rightarrow J/\psi\eta_{c}}&=&\mathcal{M}_{e}+\mathcal{M}_{f},
\end{eqnarray}
 respectively.
 
With the total amplitudes defined in Eq.~(\ref{eq:16}), one can estimate the  partial width of the above three decay processes by 
\begin{eqnarray}
\Gamma_{X\to ...}&=&\frac{1}{8\pi}\frac{|\vec{p}|}{m_{X}^{2}}\left|\ \overline{\mathcal{M}_{X\to ...}}\ \right|^{2},
\end{eqnarray}
where the overline above indicates the sum over the spin of the final states, and 
$|\vec{p}|=\sqrt{(m_X^2-(m_3+m_4)^2)(m_X^2-(m_3-m_4)^2)}/(2m_X)$  is the momentum of the daughter particles in the mother particle rest frame.

\begin{figure}[t]
  \centering
  \includegraphics[width=8.0cm]{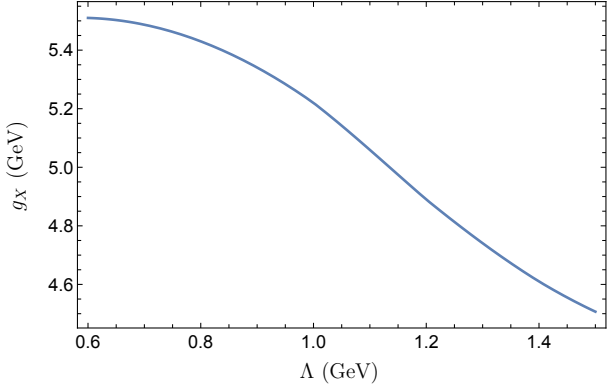}
  \caption{The coupling constant $g_X$ depending on model parameter $\Lambda$.}\label{Fig:CP}
\end{figure}

\section{Numerical results and discussions}
\label{sec:Sec4}

Before we discuss the hidden charm decay widths of $X(4014)$,  we have to clarify the values of the relevant coupling constants. In the heavy quark effective theory, the coupling constants between $S-$wave charmonia and the charmed meson pair can be related to a gauge coupling $g_{1}$ by,~\cite{Oh:2000qr,Colangelo:2002mj,Casalbuoni:1996pg}
\begin{eqnarray}
g_{\psi D^{*}D}&=&2g_{1}\sqrt{{m_{D^{*}}{m_{D}}}/m_{\psi}},\nonumber\\
g_{\psi D^{*}D^{*}}&=&2g_{1}\sqrt{m_{\psi}}m_{D^{*}},\nonumber\\
g_{\eta_{c}D^{*}D}&=&2g_{1}\sqrt{m_{D}m_{D^{*}}m_{\eta_{c}}},\nonumber\\
g_{\eta_{c}D^{*}D^{*}}&=&2g_{1}m_{D^{*}}/\sqrt{m_{\eta_{c}}},
\end{eqnarray}
where $g_{1}=\sqrt{m_{\psi}}/({2m_{D}f_{\psi}})$ and  $f_{\psi}=426 \mathrm{MeV}$ being the decay constant of  $J/\psi$ meson~\cite{Colangelo:2002mj}. Considering the heavy quark limit and chiral symmetry, the coupling constants between the light meson and charmed meson pair have the following relationship~\cite{Chen:2019asm,Liu:2011xc,Isola:2003fh,Falk:1992cx}
\begin{eqnarray}
g_{DD\mathcal{V}}&=&g_{D^{*}D^{*}\mathcal{V}}=\frac{\beta g_{V}}{\sqrt{2}},\nonumber\\
f_{D^{*}D\mathcal{V}}&=&\frac{f_{D^{*}D^{*}\mathcal{V}}}{m_{D^{*}}}=\frac{\lambda g_V}{\sqrt{2}} , \, \nonumber\\
g_{D^{*}D\mathcal{P}}&=&\frac{2g}{f_{\pi}}\sqrt{m_{D}m_{D^{*}}},\nonumber\\
g_{D^{*}D^{*}\mathcal{P}}&=&\frac{g_{D^{*}D\mathcal{P}}}{\sqrt{m_{D}m_{D^{*}}}},
\end{eqnarray}
where the parameter $g_V = {m_\rho /f_\pi}$ with $f_\pi = 132$ MeV being the pion decay constant and $\beta=0.9$ ~\cite{Casalbuoni:1996pg}. By matching the form factor obtains from the light cone sum rule with that calculated from lattice QCD, one obtained the parameters $\lambda = 0.56 \, {\rm GeV}^{-1} $ and $g=0.59$~\cite{Isola:2003fh}.

\begin{figure}[t]
  \centering
  \includegraphics[width=8.0cm]{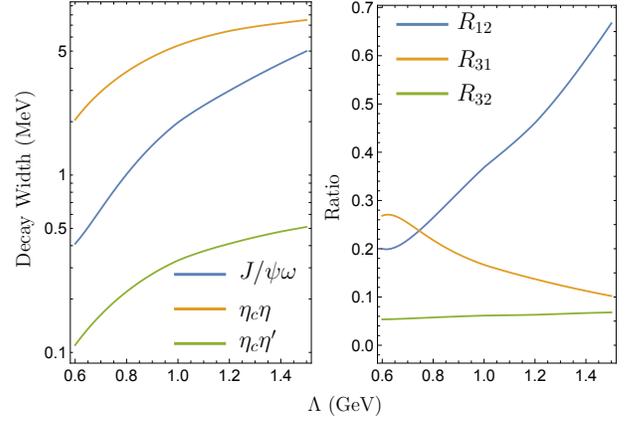}
  \caption{The $\Lambda$ dependences of the hidden charm decay widths (left panel) and their ratios (right panel).}\label{Fig:DW}
\end{figure}

As for the coupling constants $g_X$, it can be estimated by the compositeness condition as given in Eq.~(\ref{Eq:ComZ}). In the present work, a model parameter $\Lambda$ is introduced in the correlation function. Empirically, $\Lambda$ should be of the order of 1 GeV \cite{Faessler:2007gv, Faessler:2007us,Chen:2015igx,Xiao:2020ltm}. It should be clarified that the parameter $\Lambda$ can not be determined by the first principle.  Its value, in an alternative way, is usually fixed by comparing the theoretical estimations with the corresponding experimental measurements. Unfortunately, the present experimental data for $X(4014)$ is not abundant to determine the accurate value of $\Lambda$. Thus in the present work, we varies $\Lambda$ in a sizable range around 1 GeV, which is from 0.6 to 1.5 GeV. The coupling constant $g_X$ depending on model parameter $\Lambda$ is present in Fig.~\ref{Fig:CP}. From the figure one can find that the coupling constant $g_X$ decreases with the increasing of $\Lambda$. In particular, when $\Lambda$ increases from 0.6 to 1.5 GeV, the coupling constants $g_{X}$ decreases from 5.51 to 4.50 GeV.

The estimated partial widths of the $X(4014)\to J/\psi\omega$, $\eta_{c}\eta$ and $ \eta_{c}\eta^{\prime}$ are presented in Fig.~\ref{Fig:DW}. Our estimations indicate that the partial widths of all these hidden charm decay processes increase with the increasing of $\Lambda$. Particularly, in the consider parameter range the partial widths of the hidden charm decays are,
\begin{eqnarray}
\Gamma(X\to J/\psi\omega)&=&(0.41\sim 5.00)\ \mathrm{MeV},\nonumber\\
\Gamma(X\to \eta_{c}\eta)&=&(2.05\sim 7.49)\  \mathrm{MeV},\nonumber\\
\Gamma(X\to \eta_{c}\eta^{\prime})&=&(0.11\sim 0.51)\ \mathrm{MeV},
\end{eqnarray}  
respectively. As for $X(4014)$, the  measured upper limit of the width is about 16.5 MeV. Our estimations in the present work indicate that the total widths of these three channels can reach up to 13 MeV, which is still safely below the upper limit of the width of $X(4014)$. Moreover, the $\Lambda$ dependence of the partial widths are very similar, thus, one can check the ratios of these widths, here we define the ratios as $R_{12}= \Gamma_{X\to J/\psi \omega}/ \Gamma_{X\to \eta_c \eta}$, $R_{31}= \Gamma_{X\to \eta_c  \eta^\prime}/ \Gamma_{X\to J/\psi \omega }$ and $R_{32}= \Gamma_{X\to \eta_c  \eta^\prime}/ \Gamma_{X\to \eta_c \eta}$. In the right panel of Fig.~\ref{Fig:DW}, we present the $\Lambda$ dependence of these ratios. Our estimations indicate that in the consider $\Lambda$ range, one has $0.20<R_{12}<0.67$, $0.10<R_{31}<0.27$ and $0.05<R_{32}<0.07$, respectively. 

\begin{figure}[t]
  \centering
  \includegraphics[width=8.0cm]{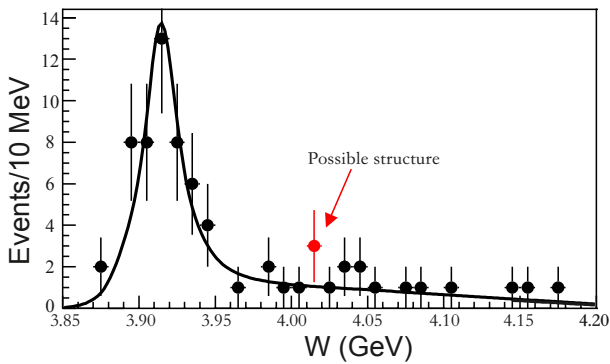}
  \caption{The measured cross sections (points with error) and fit result (solid curve) for $\gamma \gamma \to J/\psi \omega$ reported by Belle Collaboration~\cite{Belle:2009and}.}\label{Fig:Jpsiomega}
\end{figure}

Considering the fact that the $X(4014)$ is observed in the $\gamma \gamma \to \psi(2S)\gamma$ process, and the present estimations indicate that the width of $X(4014) \to J/\psi \omega$ is sizable. Thus one can expect to observe $X(4140)$ in the $\gamma \gamma \to J/\psi \omega $ process. In the year of 2009, the Belle Collaboration reported the cross sections for $\gamma \gamma \to J/\psi \omega$, where a new charmonium-like state, named $X(3915)$ was observed~ \cite{Belle:2009and}. From the experimental data as shown in Fig.~\ref{Fig:Jpsiomega}, one can find there are a number of events at $\sqrt{s}=4.015$ GeV, and it seems that there is structure corresponding to $X(4014)$, which may be checked by further analysis with a larger data sample. 

\section{Summary}
\label{sec:Sec5}
Stimulated by the recent observation of the new structure, named $X(4014)$,  in the process $\gamma \gamma \to \gamma \psi(2S)$ by the Belle Collaboration, we evaluate the possibility of interpreting $X(4014)$ as a $D^\ast \bar{D}^\ast$ molecular state. In Ref. \cite{Duan:2022upr}, we find the most possible $I(J^{PC})$ quantum numbers of $X(4104)$ are $0(0^{++})$. In the present work, we further checked such possibility by investigating the hidden charm decay properties of $X(4014)$ in the $D^\ast \bar{D}^\ast$ molecular scenario. Here, three hidden charm channels are considered, which are $J/\psi \omega$, $\eta_c \eta$ and $\eta_c \eta^\prime$, respectively. Our estimations indicates that the partial width of $\eta_c \eta$ is much larger than the other two channels and the one of $\eta_c \eta$ is also sizable in the consider parameter range. 

Considering that $X(4014)$ is observed in $\gamma \gamma $ collision process, and the present estimations indicate the width of $X(4014) \to J/\psi \omega $ is sizable, we suggest to search $X(4014)$ in the $\gamma \gamma \to J/\psi \omega $ process, which should be accessible in Belle II.

\section{ACKNOWLEDGMENTS} This work is supported by the National Natural Science Foundation of China under the Grant Nos. 11775050, 12175037, and ~11475192, the Sino-German CRC 110 "Symmetries and the Emergence of Structure in QCD" project by NSFC under the Grant
No.~12070131001, the Key Research Program of Frontier Sciences, CAS, under the Grant No.~Y7292610K1,
and the National Key Research and Development Program of China under Contracts No. 2020YFA0406300.
Supports from IHEP Innovation Fund under the grant No. Y4545190Y2 is also appreciated.


\begin{thebibliography}{00}

\bibitem{Weinberg:1965zz}
S.~Weinberg,
Phys. Rev. \textbf{137} (1965), B672-B678
doi:10.1103/PhysRev.137.B672

\bibitem{Xiao:2016hoa}
C.~J.~Xiao, D.~Y.~Chen and Y.~L.~Ma,
Phys. Rev. D \textbf{93} (2016) no.9, 094011
doi:10.1103/PhysRevD.93.094011
[arXiv:1601.06399 [hep-ph]].

\bibitem{Dong:2017gaw}
Y.~Dong, A.~Faessler and V.~E.~Lyubovitskij,
Prog. Part. Nucl. Phys. \textbf{94} (2017), 282-310
doi:10.1016/j.ppnp.2017.01.002

\bibitem{Faessler:2007gv}
A.~Faessler, T.~Gutsche, V.~E.~Lyubovitskij and Y.~L.~Ma,
Phys. Rev. D \textbf{76} (2007), 014005
doi:10.1103/PhysRevD.76.014005
[arXiv:0705.0254 [hep-ph]].

\bibitem{Faessler:2007us}
A.~Faessler, T.~Gutsche, V.~E.~Lyubovitskij and Y.~L.~Ma,
Phys. Rev. D \textbf{76} (2007), 114008
doi:10.1103/PhysRevD.76.114008
[arXiv:0709.3946 [hep-ph]].

\bibitem{Wang:2020dgr}
Z.~G.~Wang,
Int. J. Mod. Phys. A \textbf{36} (2021) no.15, 2150107
doi:10.1142/S0217751X21501074
[arXiv:2012.11869 [hep-ph]].

\bibitem{Lebed:2022vks}
R.~F.~Lebed and S.~R.~Martinez,
[arXiv:2207.01101 [hep-ph]].

\bibitem{Belle:2003nnu}
S.~K.~Choi \textit{et al.} [Belle],
Phys. Rev. Lett. \textbf{91} (2003), 262001
doi:10.1103/PhysRevLett.91.262001
[arXiv:hep-ex/0309032 [hep-ex]].

\bibitem{Liu:2008fh}
Y.~R.~Liu, X.~Liu, W.~Z.~Deng and S.~L.~Zhu,
Eur. Phys. J. C \textbf{56} (2008), 63-73
doi:10.1140/epjc/s10052-008-0640-4
[arXiv:0801.3540 [hep-ph]].

\bibitem{Wu:2021udi}
Q.~Wu, D.~Y.~Chen and T.~Matsuki,
Eur. Phys. J. C \textbf{81} (2021) no.2, 193
doi:10.1140/epjc/s10052-021-08984-2
[arXiv:2102.08637 [hep-ph]].

\bibitem{Voloshin:1976ap}
M.~B.~Voloshin and L.~B.~Okun,
JETP Lett. \textbf{23} (1976), 333-336

\bibitem{DeRujula:1976zlg}
A.~De Rujula, H.~Georgi and S.~L.~Glashow,
Phys. Rev. Lett. \textbf{38} (1977), 317
doi:10.1103/PhysRevLett.38.317

\bibitem{Tornqvist:1993ng}
N.~A.~Tornqvist,
Z. Phys. C \textbf{61} (1994), 525-537
doi:10.1007/BF01413192
[arXiv:hep-ph/9310247 [hep-ph]].

\bibitem{Tornqvist:2004qy}
N.~A.~Tornqvist,
Phys. Lett. B \textbf{590} (2004), 209-215
doi:10.1016/j.physletb.2004.03.077
[arXiv:hep-ph/0402237 [hep-ph]].

\bibitem{Wu:2019vbk}
Q.~Wu, D.~Y.~Chen, X.~J.~Fan and G.~Li,
Eur. Phys. J. C \textbf{79} (2019) no.3, 265
doi:10.1140/epjc/s10052-019-6784-6
[arXiv:1902.05737 [hep-ph]].

\bibitem{Dong:2013iqa}
Y.~Dong, A.~Faessler, T.~Gutsche and V.~E.~Lyubovitskij,
Phys. Rev. D \textbf{88} (2013) no.1, 014030
doi:10.1103/PhysRevD.88.014030
[arXiv:1306.0824 [hep-ph]].

\bibitem{Liu:2009qhy}
X.~Liu, Z.~G.~Luo, Y.~R.~Liu and S.~L.~Zhu,
Eur. Phys. J. C \textbf{61} (2009), 411-428
doi:10.1140/epjc/s10052-009-1020-4
[arXiv:0808.0073 [hep-ph]].

\bibitem{Sun:2012zzd}
Z.~F.~Sun, Z.~G.~Luo, J.~He, X.~Liu and S.~L.~Zhu,
Chin. Phys. C \textbf{36} (2012), 194-204
doi:10.1088/1674-1137/36/3/002

\bibitem{Li:2012ss}
N.~Li, Z.~F.~Sun, X.~Liu and S.~L.~Zhu,
Phys. Rev. D \textbf{88} (2013) no.11, 114008
doi:10.1103/PhysRevD.88.114008
[arXiv:1211.5007 [hep-ph]].

\bibitem{Xu:2017tsr}
H.~Xu, B.~Wang, Z.~W.~Liu and X.~Liu,
Phys. Rev. D \textbf{99} (2019) no.1, 014027
[erratum: Phys. Rev. D \textbf{104} (2021) no.11, 119903]
doi:10.1103/PhysRevD.99.014027
[arXiv:1708.06918 [hep-ph]].

\bibitem{Liu:2019stu}
M.~Z.~Liu, T.~W.~Wu, M.~Pavon Valderrama, J.~J.~Xie and L.~S.~Geng,
Phys. Rev. D \textbf{99} (2019) no.9, 094018
doi:10.1103/PhysRevD.99.094018
[arXiv:1902.03044 [hep-ph]].

\bibitem{Ding:2020dio}
Z.~M.~Ding, H.~Y.~Jiang and J.~He,
Eur. Phys. J. C \textbf{80} (2020) no.12, 1179
doi:10.1140/epjc/s10052-020-08754-6
[arXiv:2011.04980 [hep-ph]].

\bibitem{Li:2021zbw}
N.~Li, Z.~F.~Sun, X.~Liu and S.~L.~Zhu,
Chin. Phys. Lett. \textbf{38} (2021) no.9, 092001
doi:10.1088/0256-307X/38/9/092001
[arXiv:2107.13748 [hep-ph]].

\bibitem{Chen:2021vhg}
R.~Chen, Q.~Huang, X.~Liu and S.~L.~Zhu,
Phys. Rev. D \textbf{104} (2021) no.11, 114042
doi:10.1103/PhysRevD.104.114042
[arXiv:2108.01911 [hep-ph]].

\bibitem{Agaev:2021vur}
S.~S.~Agaev, K.~Azizi and H.~Sundu,
Nucl. Phys. B \textbf{975} (2022), 115650
doi:10.1016/j.nuclphysb.2022.115650
[arXiv:2108.00188 [hep-ph]].

\bibitem{Ren:2021dsi}
H.~Ren, F.~Wu and R.~Zhu,
Adv. High Energy Phys. \textbf{2022} (2022), 9103031
doi:10.1155/2022/9103031
[arXiv:2109.02531 [hep-ph]].

\bibitem{Albaladejo:2021vln}
M.~Albaladejo,
Phys. Lett. B \textbf{829} (2022), 137052
doi:10.1016/j.physletb.2022.137052
[arXiv:2110.02944 [hep-ph]].

\bibitem{Dong:2021bvy}
X.~K.~Dong, F.~K.~Guo and B.~S.~Zou,
Commun. Theor. Phys. \textbf{73} (2021) no.12, 125201
doi:10.1088/1572-9494/ac27a2
[arXiv:2108.02673 [hep-ph]].

\bibitem{Baru:2021ldu}
V.~Baru, X.~K.~Dong, M.~L.~Du, A.~Filin, F.~K.~Guo, C.~Hanhart, A.~Nefediev, J.~Nieves and Q.~Wang,
[arXiv:2110.07484 [hep-ph]].

\bibitem{Du:2021zzh}
M.~L.~Du, V.~Baru, X.~K.~Dong, A.~Filin, F.~K.~Guo, C.~Hanhart, A.~Nefediev, J.~Nieves and Q.~Wang,
Phys. Rev. D \textbf{105} (2022) no.1, 014024
doi:10.1103/PhysRevD.105.014024
[arXiv:2110.13765 [hep-ph]].



\bibitem{BESIII:2013ris}
M.~Ablikim \textit{et al.} [BESIII],
Phys. Rev. Lett. \textbf{110} (2013), 252001
doi:10.1103/PhysRevLett.110.252001
[arXiv:1303.5949 [hep-ex]].

\bibitem{Belle:2013yex}
Z.~Q.~Liu \textit{et al.} [Belle],
Phys. Rev. Lett. \textbf{110} (2013), 252002
[erratum: Phys. Rev. Lett. \textbf{111} (2013), 019901]
doi:10.1103/PhysRevLett.110.252002
[arXiv:1304.0121 [hep-ex]].

\bibitem{Xiao:2013iha}
T.~Xiao, S.~Dobbs, A.~Tomaradze and K.~K.~Seth,
Phys. Lett. B \textbf{727} (2013), 366-370
doi:10.1016/j.physletb.2013.10.041
[arXiv:1304.3036 [hep-ex]].

\bibitem{LHCb:2021auc}
R.~Aaij \textit{et al.} [LHCb],
Nature Commun. \textbf{13} (2022) no.1, 3351
doi:10.1038/s41467-022-30206-w
[arXiv:2109.01056 [hep-ex]].

\bibitem{LHCb:2021vvq}
R.~Aaij \textit{et al.} [LHCb],
doi:10.1038/s41567-022-01614-y
[arXiv:2109.01038 [hep-ex]].

\bibitem{He:2013nwa}
J.~He, X.~Liu, Z.~F.~Sun and S.~L.~Zhu,
Eur. Phys. J. C \textbf{73} (2013) no.11, 2635
doi:10.1140/epjc/s10052-013-2635-z
[arXiv:1308.2999 [hep-ph]].

\bibitem{Chen:2013omd}
W.~Chen, T.~G.~Steele, M.~L.~Du and S.~L.~Zhu,
Eur. Phys. J. C \textbf{74} (2014) no.2, 2773
doi:10.1140/epjc/s10052-014-2773-y
[arXiv:1308.5060 [hep-ph]].


\bibitem{BESIII:2013ouc}
M.~Ablikim \textit{et al.} [BESIII],
Phys. Rev. Lett. \textbf{111} (2013) no.24, 242001
doi:10.1103/PhysRevLett.111.242001
[arXiv:1309.1896 [hep-ex]].

\bibitem{Wang:2022fdu}
Z.~G.~Wang,
[arXiv:2207.00947 [hep-ph]].

\bibitem{Sakai:2021qrg}
S.~Sakai,
[arXiv:2107.11026 [hep-ph]].

\bibitem{Ortega:2018cnm}
P.~G.~Ortega, J.~Segovia, D.~R.~Entem and F.~Fern\'andez,
Eur. Phys. J. C \textbf{79} (2019) no.1, 78
doi:10.1140/epjc/s10052-019-6552-7
[arXiv:1808.00914 [hep-ph]].

\bibitem{Xiao:2019spy}
L.~Y.~Xiao, G.~J.~Wang and S.~L.~Zhu,
Phys. Rev. D \textbf{101} (2020) no.5, 054001
doi:10.1103/PhysRevD.101.054001
[arXiv:1912.12781 [hep-ph]].

\bibitem{Ke:2016owt}
H.~W.~Ke and X.~Q.~Li,
Eur. Phys. J. C \textbf{76} (2016) no.6, 334
doi:10.1140/epjc/s10052-016-4183-9
[arXiv:1601.03575 [hep-ph]].

\bibitem{Belle:2021nuv}
X.~L.~Wang \textit{et al.} [Belle],
Phys. Rev. D \textbf{105} (2022) no.11, 112011
doi:10.1103/PhysRevD.105.112011
[arXiv:2105.06605 [hep-ex]].


\bibitem{Duan:2022upr}
M.~Y.~Duan, D.~Y.~Chen and E.~Wang,
[arXiv:2207.03930 [hep-ph]].



\bibitem{Liu:2005jb}
W.~Liu, C.~M.~Ko and L.~W.~Chen,
Nucl. Phys. A \textbf{765} (2006), 401-425
doi:10.1016/j.nuclphysa.2005.11.008
[arXiv:nucl-th/0505075 [nucl-th]].

\bibitem{Wu:2021ezz}
Q.~Wu and D.~Y.~Chen,
Phys. Rev. D \textbf{104} (2021) no.7, 074011
doi:10.1103/PhysRevD.104.074011
[arXiv:2108.06700 [hep-ph]].

\bibitem{Wu:2021cyc}
Q.~Wu, D.~Y.~Chen, W.~H.~Qin and G.~Li,
Eur. Phys. J. C \textbf{82} (2022) no.6, 520
doi:10.1140/epjc/s10052-022-10465-z
[arXiv:2111.13347 [hep-ph]].

\bibitem{Chen:2016ncs}
D.~Y.~Chen and C.~J.~Xiao,
Nucl. Phys. A \textbf{947} (2016), 26-37
doi:10.1016/j.nuclphysa.2015.12.003



\bibitem{Hayashi:1967bjx}
K.~Hayashi, M.~Hirayama, T.~Muta, N.~Seto and T.~Shirafuji,
Fortsch. Phys. \textbf{15} (1967) no.10, 625-660
doi:10.1002/prop.19670151002


\bibitem{Weinberg:1962hj}
S.~Weinberg,
Phys. Rev. \textbf{130} (1963), 776-783
doi:10.1103/PhysRev.130.776

\bibitem{Salam:1962ap}
A.~Salam,
Nuovo Cim. \textbf{25} (1962), 224-227
doi:10.1007/BF02733330

\bibitem{Casalbuoni:1996pg}
R.~Casalbuoni, A.~Deandrea, N.~Di Bartolomeo, R.~Gatto, F.~Feruglio and G.~Nardulli,
Phys. Rept. \textbf{281} (1997), 145-238
doi:10.1016/S0370-1573(96)00027-0
[arXiv:hep-ph/9605342 [hep-ph]].

\bibitem{Oh:2000qr}
Y.~s.~Oh, T.~Song and S.~H.~Lee,
Phys. Rev. C \textbf{63} (2001), 034901
doi:10.1103/PhysRevC.63.034901
[arXiv:nucl-th/0010064 [nucl-th]].

\bibitem{Colangelo:2002mj}
P.~Colangelo, F.~De Fazio and T.~N.~Pham,
Phys. Lett. B \textbf{542} (2002), 71-79
doi:10.1016/S0370-2693(02)02306-7
[arXiv:hep-ph/0207061 [hep-ph]].

\bibitem{Neubert:2005mu}
M.~Neubert,
doi:10.1142/9789812773579\_0004
[arXiv:hep-ph/0512222 [hep-ph]].

\bibitem{Kaplan:2005es}
D.~B.~Kaplan,
[arXiv:nucl-th/0510023 [nucl-th]].

\bibitem{Kaymakcalan:1983qq}
O.~Kaymakcalan, S.~Rajeev and J.~Schechter,
Phys. Rev. D \textbf{30} (1984), 594
doi:10.1103/PhysRevD.30.594


\bibitem{MARK-III:1988crp}
D.~Coffman \textit{et al.} [MARK-III],
Phys. Rev. D \textbf{38} (1988), 2695
[erratum: Phys. Rev. D \textbf{40} (1989), 3788]
doi:10.1103/PhysRevD.38.2695

\bibitem{DM2:1988bfq}
J.~Jousset \textit{et al.} [DM2],
Phys. Rev. D \textbf{41} (1990), 1389
doi:10.1103/PhysRevD.41.1389

\bibitem{Chen:2019asm}
R.~Chen, Z.~F.~Sun, X.~Liu and S.~L.~Zhu,
Phys. Rev. D \textbf{100} (2019) no.1, 011502
doi:10.1103/PhysRevD.100.011502
[arXiv:1903.11013 [hep-ph]].

\bibitem{Liu:2011xc}
Y.~R.~Liu and M.~Oka,
Phys. Rev. D \textbf{85} (2012), 014015
doi:10.1103/PhysRevD.85.014015
[arXiv:1103.4624 [hep-ph]].

\bibitem{Isola:2003fh}
C.~Isola, M.~Ladisa, G.~Nardulli and P.~Santorelli,
Phys. Rev. D \textbf{68} (2003), 114001
doi:10.1103/PhysRevD.68.114001
[arXiv:hep-ph/0307367 [hep-ph]].

\bibitem{Falk:1992cx}
A.~F.~Falk and M.~E.~Luke,
Phys. Lett. B \textbf{292} (1992), 119-127
doi:10.1016/0370-2693(92)90618-E
[arXiv:hep-ph/9206241 [hep-ph]].


\bibitem{Chen:2015igx}
D.~Y.~Chen and Y.~B.~Dong,
Phys. Rev. D \textbf{93} (2016) no.1, 014003
doi:10.1103/PhysRevD.93.014003
[arXiv:1510.00829 [hep-ph]].

\bibitem{Xiao:2020ltm}
C.~J.~Xiao, D.~Y.~Chen, Y.~B.~Dong and G.~W.~Meng,
Phys. Rev. D \textbf{103} (2021) no.3, 034004
doi:10.1103/PhysRevD.103.034004
[arXiv:2009.14538 [hep-ph]].

\bibitem{Belle:2009and}
S.~Uehara \textit{et al.} [Belle],
Phys. Rev. Lett. \textbf{104} (2010), 092001
doi:10.1103/PhysRevLett.104.092001
[arXiv:0912.4451 [hep-ex]].
Copy to ClipboardDownload


\end{thebibliography}
\end{document}